\documentclass [a4paper]{scrartcl}
\usepackage [T1]{fontenc}
\usepackage [utf8]{inputenc}
\usepackage [english] {babel}
\usepackage{graphicx}
\usepackage{subfig}
\usepackage{authblk}

\title{ $H\rightarrow \mu^+ \mu^- $ at ILC}
\subtitle {Talk presented at the International Workshop on Future Linear Colliders (LCWS15), Whistler, Canada, 2-6 November 2015}

\author[1]{Michele Faucci Giannelli}
\author[2]{Sara Celani}
\affil[1]{Department of Physics, Royal Holloway University of London, Surrey, United Kingdom}
\affil[2]{Dipartimento di Fisica, Sapienza Università di Roma, Roma, Italy }

\begin{document}
\maketitle
\section{Introduction}
The discovery of a Higgs boson with a mass of 125\,GeV at the LHC~\cite{CMS:Higgs, ATLAS:Higgs} has enhanced the physics case for the construction of the International Linear Collider (ILC)~\cite{ILC} since this new machine will allow measuring the properties of the new particle with a precision unreachable at the LHC. The most crucial properties to measure are the couplings to other particles since many Beyond Standard Model scenarios predict small variations, of the order of few percent,  with respect to the Standard Model. Hence a high precision is required to discriminate between models. Although it is relatively easy to  measure the couplings to third generation particles and to bosons, due to their large mass and therefore coupling, it is very difficult to measure coupling to second generation particles. The analysis we describe aims at filling this gap by measuring the coupling of the Higgs boson to the muons. The presence of the two muons from the Higgs decay in signal events makes this channel an ideal benchmark for measuring the performance of the tracking system of the International Large Detector (ILD)~\cite{ILD}.

We focussed on the analysis of the $\nu_e \bar{\nu_e} H, H\rightarrow \mu^+ \mu^- $ channel at centre mass energy of 1\,TeV. At this energy the WW-fusion process (Fig 1a) has a larger cross section than the Higgsstrahlung (Fig 1b), so much that the latter can be neglected. We assumed an integrated luminosity of $500\,fb^{-1}$ and polarised beams $(P_{e^-}, P_{e^+}) = (-0.8 , +0.2)$. We considered the finale states with $\nu_e \bar{\nu}_e \mu^+ \mu^- $, where the two muons were not from an Higgs decay, as the main background processes. We developed a new analysis starting from the one published for the Detector Baseline Document (DBD)~\cite{Tino:Note}.

\begin{figure}
\centering
\subfloat[][\emph{WW-fusion}.]
	{\includegraphics[scale=0.7]{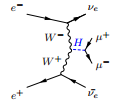}} \quad
\subfloat[][\emph{Higgsstrahlung}]
	{\includegraphics[scale=0.7]{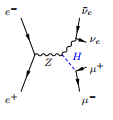}}\quad
\caption{The two processes for Higgs production.}
\label{fig:subfig}
\end{figure}

\section {Software and Monte Carlo Samples}

We used the Monte Carlo samples produced by the ILD detector optimisation group for the DBD.
The event generation was performed using WHIZARD v1.95~\cite{Kilian:2007gr, Moretti:2001zz}. 
PYTHIA~\cite{PYTHIA6} was used for the modelling the fragmentation while the decays of $\tau$ leptons were handled by TAUOLA~\cite{TAUOLA}. 
The simulation of the ILD detector was carried out with GEANT4~\cite{GEANT4}. The event
reconstruction was performed using ILCSOFT v01-16~\cite{ILCSOFT} framework while the analysis used ROOT 
v5.34.25~\cite{ROOT} and the TMVA~\cite{TMVA}.

Given the small branching ratio of the Higgs to muon, a dedicated production was generated forcing the Higgs to decay only to muons in the channel, $\nu_e \bar{\nu}_e H, H \rightarrow \nu_e \bar{\nu}_e \mu^+ \mu^- $. For the backgrounds we only considered those channels that were found to be significant in~\cite{Tino:Note}: $\nu_e \bar{\nu}_e Z \rightarrow \nu_e \bar{\nu}_e l^+l^-$, the single boson Z leptonic decay, $WW/ZZ \rightarrow \nu_l \bar{\nu}_l l^+ l^-$, double boson W or Z leptonic decay, $ WW \rightarrow \nu_{\mu} l^+ \bar{\nu_{\tau}} l^-$, WW scattering, $\gamma \gamma \rightarrow \nu_e \bar{\nu}_e l^+l^-$, the machine background. The processes are summarised in Fig 2 with their respective cross sections.

\begin{figure}
\centering
\includegraphics[scale=0.3]{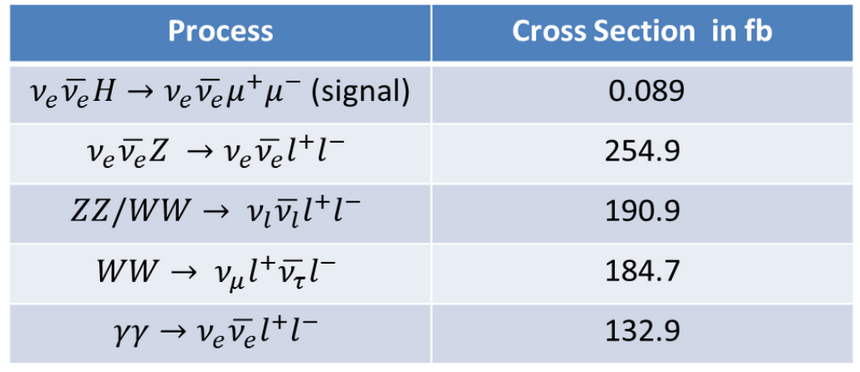}
\caption{The signal and background processes are shown with their respective cross sections.}
\end{figure}

\section {Muons identification strategies}

The core of the analysis is the identification of the two muons coming from the Higgs boson decay. Since the publication of the DBD, the ILC community improved the particles identification which allowed us to use new strategies. We compared three methods to identify a particle as a muon produced by a Higgs decay. The first method is used as a baseline and is identical to the one used in the DBD~\cite{Tino:Note}, this is a simple cut selection based only on calorimeter information; the second method is an improvement of the first selection method as it also uses the muon tracker; finally, the third method is based on a Neural Network as implemented in the Isolated Lepton Tagging Processor (July version)~\cite{Junping:Iso}.

The requests to identify a muon for both first and second method, are summarised in Fig 3. For each muon selection method, the following analysis was performed.

\begin{figure}
\centering
\includegraphics[scale=0.3] {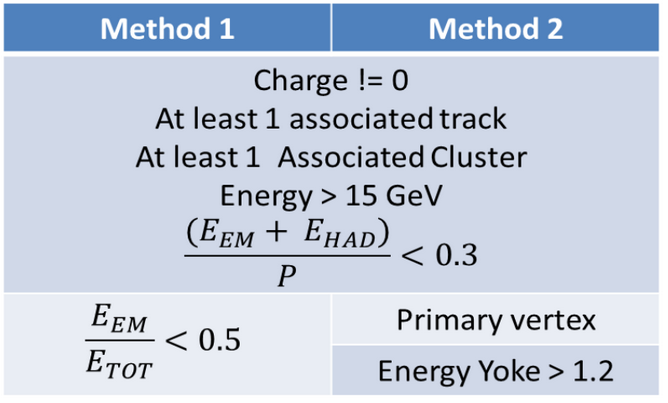}
\caption{ Features that a particle must have to be recognized as a muon both for the first method (on the left) and for the second method (on the right). Most of them are in common in the two methods. }
\end{figure}

\section{Analysis strategy}

In addition to the requirement of identifying two muons as described above, a series of cuts was implemented to reject many backgrounds. These cuts are similar to those described in~\cite{Tino:Note} to allow a direct comparison of the results.

\begin{itemize}
\item A cut on the invariant mass of the two muons: $|M_{ll}-125|<30$\,GeV.
\item A cut on the angle between muons $\theta_{ll}>0.5$\,rad. The angle distribution is shown in Fig 4a; it is clear that only background events have an angle smaller than 0.5\,rad.
\item A cut on the missing energy $E_{miss}>300$\,GeV. This quantity is expected to be larger for signal events than for most background as shown in Fig 4b.
\item A cut on the number of charged particles (<4) and on the number of charged leptons (<3), as no additional charged particles are expected in signal events. 
\end {itemize}

\begin{figure}
\centering
\subfloat[][\emph{Angle between muons}.]
	{\includegraphics[scale=0.28]{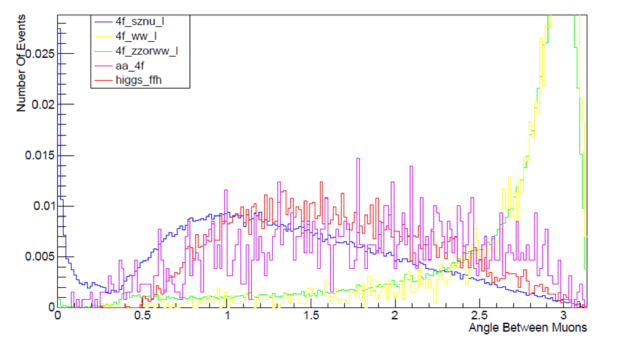}} 
\subfloat[][\emph{Missing energy}]
	{\includegraphics[scale=0.28]{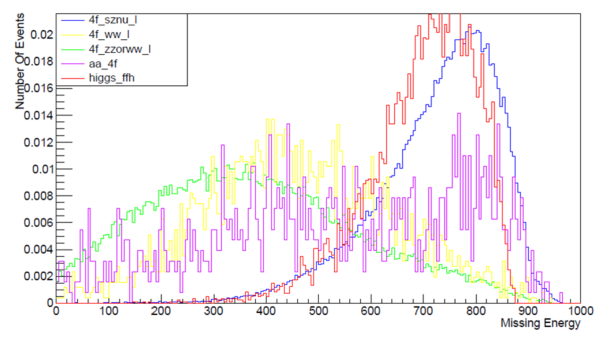}}
\caption{Distribution of the quantities used to implement cuts; angle between muons (a) and missing energy (b).}
\label{fig:subfig}
\end{figure}

The table shown in Fig 5 shows the number of events passing each cut for every channel. The final column shows the number of selected events normalised to $500\,fb^{-1}$. A similar table was produced for each muon identification strategy; for simplicity only the second selection is shown, with the other selections being very similar.

Using these cuts we calculated the significance $\sigma$, defined as 
\[ \frac{S}{\sqrt{S+B}} \]

where S is the number of signal events and B is the number of background events. Without further cuts or optimisations the significance is $\sigma \approx 0.54$. 
This poor result is expected since, at this stage, we have not yet used the strongest discriminating variable, the Higgs invariant mass. It is worth highlighting the large number of simulated events that pass the selection as shown in the second column from the right in Fig 5. Because of this high statistics it was possible to perform a multivariate analysis. 

\begin{figure}
\centering
\includegraphics[scale=0.55] {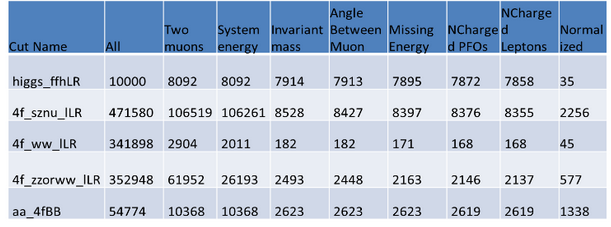}
\caption{Number of the signal and the background events after all the cuts for the second muon identification method.}
\end{figure}

\section{Multivariate Analysis}

For each muon identification methods we compared two multivariate methods implemented in the TMVA.
We focussed on the Fisher and the Artificial Neural Network (ANN) methods. The variables used as inputs were: 

\begin{itemize}
\item the angle between muons;
\item the leading muon energy;
\item the subleading muon energy; 
\item the missing energy;
\item the di-muon system energy;
\item the leading muon $cos\theta$, where $\theta$ is the angle between muon and the beam line;
\item the subleading muon $cos\theta$;
\item the di-muon system invariant mass;
\end{itemize}

\begin{figure}
\centering
\subfloat[][]
	{\includegraphics[scale=0.4]{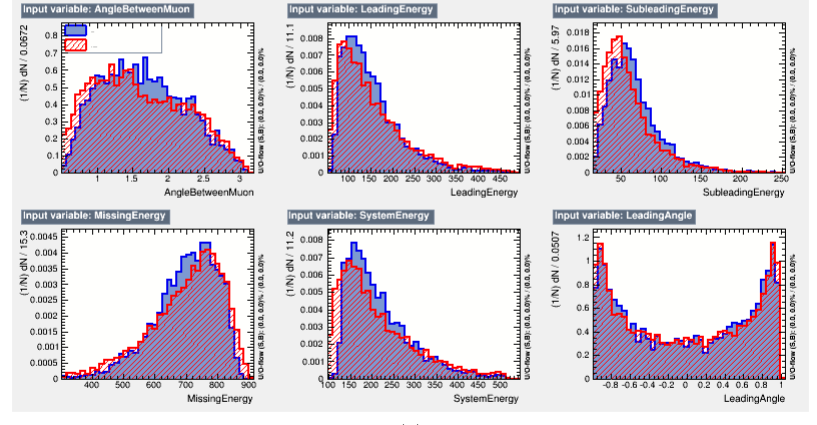}} \\
\subfloat[][]
	{\includegraphics[scale=0.4]{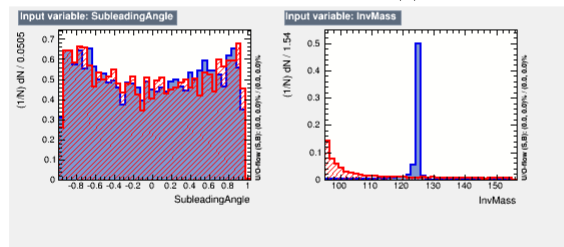}}
\caption{Distribution of the TMVA input variables. Blue are signal events, red are the sum of all background events.}
\label{fig:subfig}
\end{figure}

As shown in Figure 7, a better separation was achieved for the ANN method in comparison to the Fisher method. 

\begin{figure}
\centering
\includegraphics[scale=0.45] {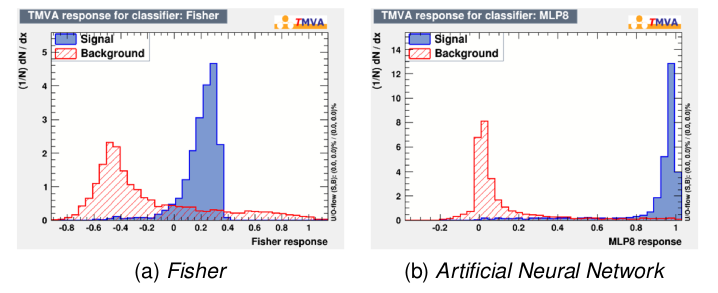}
\caption {Separation for Fisher and Artificial Neural Network.  Blue are signal events, red are the sum of all background events.}
\end{figure}

\section{Results}

\begin{figure}
\centering
\subfloat[][ROC curve]
	{\includegraphics[scale=0.5]{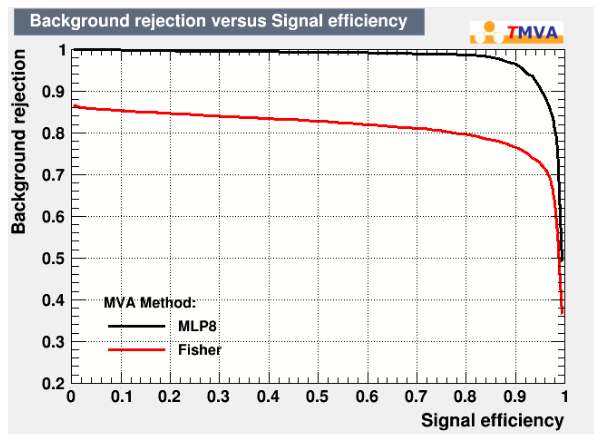}} \\
\subfloat[][Significance]
	{\includegraphics[scale=0.5]{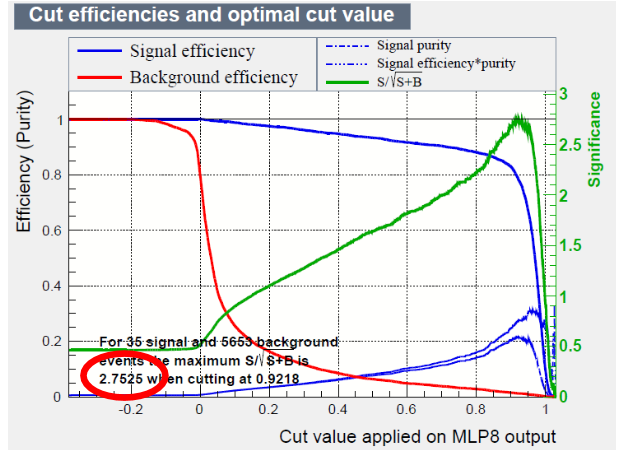}}
\caption{The ROC curves for both Fisher and ANN methods (a) and the significance for the ANN (b).}
\label{fig:subfig}
\end{figure}

Each muon identification method was trained using the full amount of selected events by that strategy to avoid any bias. The results for the second muon identification method are summarised in Fig 8. The main result is shown in Fig 8b. The significance for this selection is $\sigma \approx 2.75$. This is a significant improvement with respect to the $\sigma \approx 2.3$ achieved in~\cite{Tino:Note}.

The significances obtained for the three muon selection strategies are shown in the table of Fig 9. TMVA applied to the selection used in~\cite{Tino:Note} improved the sensitivity from 2.3 to 2.57. The better muon selection provided by the second method, combined with the neural network, allowed to reach a sensitivity of 2.75.

\begin{figure}
\centering
\includegraphics[scale=0.5] {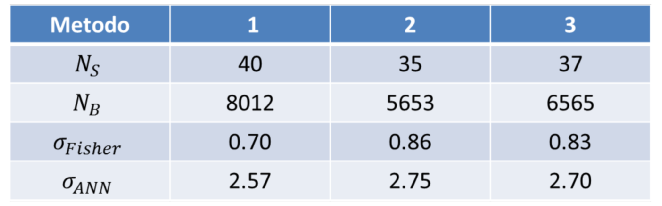}
\caption {$N_s$ number of signal events, $N_b$ number of background events passing ANN cut. All values are normalised to $500fb^{-1}$.}
\end{figure}
We also performed a study on the value of the cut on the mass window of the invariant of the two muons. Reducing the allowed window also reduces the sensitivity for the neural network. This is due to the reduced statistics for both signal and background which limits the training capabilities of the ANN. The results of the study are shown in the table of Fig 10.  

\begin{figure}
\centering
\includegraphics[scale=0.7] {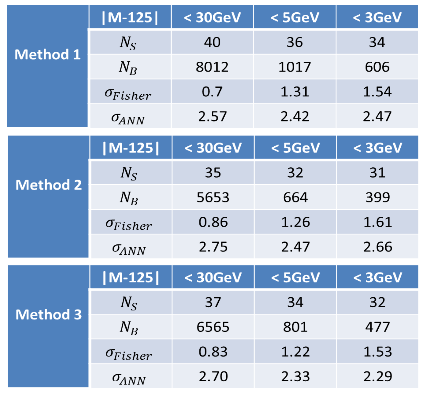}
\caption {Significance for Fisher and ANN with a different cut on the Higgs mass.}
\end{figure}

\section {Conclusion}

We performed the analysis of the $H \rightarrow \mu^+ \mu^- $ channel at 1\,TeV using the most recent samples from the ILD detector.
By using MultiVariate techniques we increased the significance with respect to the DBD result from $\sigma \approx 2.3$ to $\sigma \approx 2.57$.
Combining the neural network selection with the latest lepton reconstruction, the significance was further improved to $\sigma \approx 2.7$.

It is worth pointing out that this analysis can be further improved using the latest available Lepton Isolation Tagger which also allows the recovery of photons emitted by the muons. Further improvements to the MVA can also be implemented and we hope to include them in the near future. 

Finally we would like to stress the fact that this is not the only centre of mass at which the  $H \rightarrow \mu^+ \mu^- $ process can be studied, hence the ultimate precision of ILD to this process is expected to be higher.


\addcontentsline{toc}{section}{\refname}

\begin{thebibliography}{10}

\bibitem{CMS:Higgs}
The CMS Collaboration, “A New Boson with a Mass of 125 GeV Observed with the CMS Experiment at the Large Hadron Collider,” Science, vol. 338, pp. 1569-1575, 2012.

\bibitem{ATLAS:Higgs}
The ATLAS Collaboration, “A Massive Particle Consistent with the Standard Model Higgs Boson observed with the ATLAS Detector at the Large Hadron Collider,” Science, vol. 338, pp. 1576-1582 , 2012.

\bibitem{ILC}
Behnke, Ties et al, "The International Linear Collider Technical Design Report - Volume 1: Executive Summary''. arXiv:1306.6327 [physics.acc-ph].

\bibitem{ILD}
http://ilcild.org

\bibitem{Tino:Note}
Costantino Calancha,''Study of $ H\rightarrow \mu^+ \mu^- $ at  $\sqrt{s} = 1$\,TeV at the ILC'', LC-REP-2013-006

\bibitem{Kilian:2007gr} 
  W.~Kilian, T.~Ohl and J.~Reuter,
  `WHIZARD: Simulating Multi-Particle Processes at LHC and ILC,''
  Eur.\ Phys.\ J.\ C {\bf 71}, 1742 (2011)
  doi:10.1140/epjc/s10052-011-1742-y
  [arXiv:0708.4233 [hep-ph]].

\bibitem{Moretti:2001zz} 
  M.~Moretti, T.~Ohl and J.~Reuter, 
  ''O'Mega: An Optimizing matrix element generator,''
  hep-ph/0102195.

\bibitem{PYTHIA6} 
T. Sjostrand, S. Mrenna, and P. Z. Skands. “PYTHIA 6.4 Physics and Manual.” JHEP, vol. 05
p. 026, 2006. hep-ph/0603175.

\bibitem{TAUOLA} 
Z. Was. “TAUOLA the library for tau lepton decay, and KKMC/KORALB/KORALZ/... status
report.” Nucl. Phys. Proc. Suppl., vol. 98 pp. 96102, 2001. hep-ph/0011305.

\bibitem{GEANT4} 
S. Agostinelli et al. “Geant4 A Simulation Toolkit.” Nucl. Instrum. Methods Phys. Res., Sect. A,
vol. 506(3) pp. 250303, 2003.

\bibitem{ILCSOFT} 
http://ilcsoft.desy.de/portal/.

\bibitem{ROOT} 
R. Brun and F. Rademakers. “ROOT – an object oriented data analysis framework.” Nucl. Instrum.
Methods Phys. Res., Sect. A, vol. 389(1-2) pp. 8186, 1997.

\bibitem{TMVA} 
 A.~Hoecker, P.~Speckmayer, J.~Stelzer, 
        J.~Therhaag, E.~von Toerne, and H.~Voss,
        ``TMVA: Toolkit for Multivariate Data Analysis,''
        PoS A CAT 040 (2007) [physics/0703039].

\bibitem{Junping:Iso}
Claude Duerig, Junping Tian, ''The Isolated Lepton finder'', High Level Reconstruction Workshop, https://agenda.linearcollider.org/event/6787/session/10/contribution/17/material/slides/0.pdf

\end{thebibliography}
\end{document}